\documentclass[]{jfm}
\usepackage[latin1]{inputenc}
\usepackage[english]{babel}
\usepackage{graphicx}
\usepackage{subfig}
\usepackage{array}
\usepackage{amsmath}
\usepackage{SIunits}
\usepackage{natbib}
\graphicspath{{figspdflatex/}}

\def\dashedrule#1#2#3{{%
\dimen1=#2 \divide\dimen1 by 2
\def\@ruledash{%
\rule{\dimen1}{0pt}%
\rule[0.5ex]{#1}{0.4pt}%
\rule{\dimen1}{0pt}}%
\count1=0
\loop%
\ifnum\count1<#3%
\advance\count1 by 1%
\@ruledash%
\repeat}}

\title{Sonoluminescence and sonochemiluminescence from a microreactor}
\author[D.F. Rivas, Ashokkumar, Leong, Yasui, Tuziuti, Kentish, Lohse and Gardeniers]{D\ls A\ls V\ls I\ls D\ns F\ls E\ls R\ls N\ls A\ls N\ls D\ls E\ls Z\ns R\ls I\ls V\ls A\ls S,\ns$^1$ \thanks{Email address for correspondence: d.fernandezrivas@utwente.nl} M\ls U\ls T\ls H\ls U\ls P\ls A\ls N\ls D\ls I\ls A\ls N\ns A\ls S\ls H\ls O\ls K\ls K\ls U\ls M\ls A\ls R,\ns$^2$ T\ls H\ls O\ls M\ls A\ls S\ns L\ls E\ls O\ls N\ls G,\ns$^2$ K\ls Y\ls U\ls I\ls C\ls H\ls I\ns Y\ls A\ls S\ls U\ls I,\ns$^3$ T\ls O\ls R\ls U\ns T\ls U\ls Z\ls I\ls U\ls T\ls I,\ns$^3$ S\ls A\ls N\ls D\ls R\ls A\ns K\ls E\ls N\ls T\ls I\ls S\ls H,\ns$^2$ D\ls E\ls T\ls L\ls E\ls F\ns L\ls O\ls H\ls S\ls E\ns$^1$ \and H\ls A\ls N\ns  J.G.E. G\ls A\ls R\ls D\ls E\ls N\ls I\ls E\ls R\ls S\ns$^1$}
 
\affiliation{$^1$University of Twente, Enschede, The Netherlands
\\[\affilskip]
$^2$ School of Chemistry, Department of Chemical and Biomolecular Engineering, University of Melbourne, Australia
\\[\affilskip]
$^3$ National Institute of Advanced Industrial Science and Technology, AIST, Moriyama ku, Nagoya, Japan}

\date{\today}

\begin{document}
\maketitle


\begin{abstract}
Micromachined pits on a substrate can be used to nucleate and stabilize microbubbles in  a liquid exposed to an ultrasonic field. Under suitable conditions, the collapse of these bubbles can result in light emission (sonoluminescence, SL). Hydroxyl radicals (OH$^.$) generated during bubble collapse can react with luminol to produce light (sonochemiluminescence, SCL). SL and SCL  intensities were recorded for several regimes related to the pressure amplitude (low and high acoustic power levels) at a given ultrasonic frequency (200 kHz) for pure water, and aqueous luminol and propanol solutions. Various arrangements of pits were studied, with the number of pits ranging from no pits (comparable to a classic ultrasound reactor), to three-pits. Where there was more than one pit present, in the high pressure regime the ejected microbubbles combined into linear (two-pits) or triangular (three-pits) bubble clouds (streamers). In all situations where a pit was present on the substrate, the SL  was intensified and increased with the number of pits at both low and high power levels. For imaging SL emitting regions, Argon (Ar) saturated water was used under similar conditions. SL emission from aqueous propanol solution (50 mM) provided evidence of transient bubble cavitation. Solutions containing 0.1 mM luminol were also used to demonstrate the radical production by attaining the SCL emission regions. 
\end{abstract}


\section{Introduction}
\label{sect:intro} 

A well known effect of ultrasonic irradiation in a liquid is acoustic cavitation, which is the nucleation and consequent collapse of bubbles \citep{ashokkumar2007bubbles}. Sometimes, bubbles can cavitate in phase with the applied sound frequency. These bubbles behave as ``individual micro-reactors", as they are often accompanied by a violent collapse that leads to high pressures and temperatures within and in the local vicinity of the bubbles  \citep{lohse2005sonoluminescence,Didenko2002,bre02}. Among the events generated during such collapses, some can be identified as plasma formation, lysis of molecules yielding radicals (water molecule sonolysis is a well known example), strong pressure shockwaves, liquid jetting and  surface erosion. Although the driving condition is constant in most cases, the cavitation bubbles are never exactly the same due to the complexity of the phenomena involved \citep{lauterborn2010}, rendering it difficult to obtain a deterministic relation between driving conditions and actual bubble population. To complement direct determination of bubble size distributions which require fast imaging and intensive image processing, a good alternative is the use of indirect methods like measuring the light emitted by these bubbles.

Most researchers agree that the studies of Marinesco and Trillat \citep{marinesco1933}  and Frenzel and Shultes \citep{frenzel34} were the first in which light emission could be detected as a result of ultrasound irradiation in a liquid sample. 
After several studies to determine the actual mechanism behind this remarkable effect, sonoluminescence (SL) has been described as the light emitted by cavitation bubbles driven by an ultrasonic driving pressure field. Sonochemiluminescence (SCL) is defined in this study as the light emission when luminol reacts with OH$^{.}$ radicals.
There is a large difference between the SL emission from a single bubble (SBSL) and SL from a cluster of bubbles (MBSL, i.e. multibubble SL). For SBSL the collapse is nearly spherically symmetric and highly reproducible \citep{bre02}; and for MBSL the more frequent non repeatable asymmetric collapse produces liquid jets penetrating the hot bubble contents \citep{crum1994,flint1991,matula1997,mcnamara00}.  These differences are evident in the emission profile and spectra produced from both types of SL. The spectra collected from MBSL contain many peaks and features, whereas the spectra emanating from SBSL are normally featureless \citep{suslick1999acoustic,suslick2008inside}.
	The effect of power and frequency on bubble-size distributions of MBSL in acoustic cavitation has been studied previously using pulsed ultrasound \citep{brotchie2009}. The main conclusion was that the mean bubble size increased with increasing acoustic power and at the same time decreased with increasing ultrasound frequency. Additionally, the mean bubble size distribution of bubbles emitting SL was larger and narrower than SCL producing bubbles (centred at smaller bubble sizes and broader) meaning that the two processes result from different bubble sizes \citep{yasui2008} and the physical locations in which these bubbles exist can differ \citep{sunartio2007}. 
	MBSL has also a defined phase window of the driving pressure oscillation, meaning that in a phase window of about 30$^0$ from the full 360$^0$ SL can be detected \citep{lauterborn2010}.
Other researchers have reported singular dependences of SL and capillary pressure in small gaps with an implicit advantage in not using a light-proof box to quantify multibubble intertial cavitation thresholds \citep{Dezhkunov2004}. 
	
Sonochemistry performed in microfluidic devices has received some attention in the last decade \citep{fernandez2010,iida2004ultrasonic,lee2007influence,ohl2011sonochemistry}. In this work an extended study on an ultrasonic microreactor described before \citep{fernandez2010} is presented. The working principle of that sonoreactor is based on the
ability of small predefined crevices (pits etched in silicon substrate surface \citep{bre06,bremond2006interaction,marmottant2006microfluidics,borkent2009nucleation}) to stabilize small gas nuclei. 
When the ultrasound is turned on, a characteristic microbubble cloud appears, that would not be present in the absence of pits.
In this way a continuous locally controlled generation of cavitating microbubbles is achieved. The chemical activity of these microbubbles was previously verified by luminol SCL imaging and OH$^{.}$ radical dosimetry by using terephthalic acid \citep{fernandez2010}. 

The aim of the present work was to study the changes of SL and SCL intensities emanating from different solutions as the population of microbubbles nucleated from the pits on the silicon substrate varies. This is influenced by altering the power input to the system and the number of pits. The areas of potential application for these findings are many; to name a few we consider the ultrapurification of water for fine chemicals or pharmaceutical uses, mechanochemistry and the selective cleaning of circuit boards in which localized cavitation can avoid the damage of certain components with conventional sonication systems \citep{Cravotto2012,cobley2011initial}. Our results  can be of importance to existing non destructive testing and inspection of surfaces with localized fluorescent dye penetration which are improved by the action of localized cavitation\citep{Dezhkunov2005}.

Additionally for biological applications where a localized source of radicals, light and streaming forces are required, this system might be beneficial \citep{Ohl2003131,Dijkink2008}. 

\section{Material and Methods}    \label{sect:MatMeth}

This work focused on the measurement of the SL and SCL intensities emitted from three different systems. The first was air-saturated Milli-Q water, the second was air saturated propanol solution (50 mM) in water and the third was air saturated aqueous luminol (0.1 mM luminol in 0.1 M NaOH) solution. SL intensities were obtained for water and propanol, whereas SCL intensities were measured for luminol. Additionally SL images were recorded in argon saturated water.

In most cavitating systems, there exist populations of SL active and SCL active bubbles \citep{AshokSpatial2010}. These populations strongly overlap: SL active bubbles can be SCL active, and vice versa. SL active bubbles correspond to bubbles that satisfy suitable conditions (pressure and temperature) inside the bubble that allow for ionization and the subsequent light emission \citep{Hilgenfeldt99}. SCL active bubbles produce radicals (OH$^{.}$ radicals in this case):

\begin{equation}
H_2O \stackrel{\Delta H= 5.1 eV}\rightleftharpoons OH^{.} + H^{.} 
\label{eq:formation}
\end{equation}

Recording the light emission from luminol molecules reacting with OH$^{.}$ is a widely used method to quantify the chemical activity and map active zones in a sonoreactor. 
Evidently knowing the exact bubble size distribution and their spatial localization is very difficult (\citep{luther2001,tsochatzidis2001,fernandez2012ts}), and most studies are based on bubble dissolution when US is turned off \citep{brotchie2009,labouret2002,chen2002}. By virtue of dissolved propanol in water, information about the presence of transient cavitating bubbles can be obtained \citep{ashokkumar2009detection,price2004sonoluminescence}.  It has been shown that alcohols do not quench SL arising from transient cavitation (i.e., MBSL), as the alcohol
molecules do not have enough time to accumulate on the interface of the  transiently cavitating  bubbles \citep{price2004sonoluminescence}. That is in vast contrast to stable SBSL, where
 alcohols strongly quench the light emission \citep{toegel2000does}.

\subsection{Set-up for US experiments} \label{subsect:USSetup}
\begin{figure}
\centering
\includegraphics[width=0.6\columnwidth]{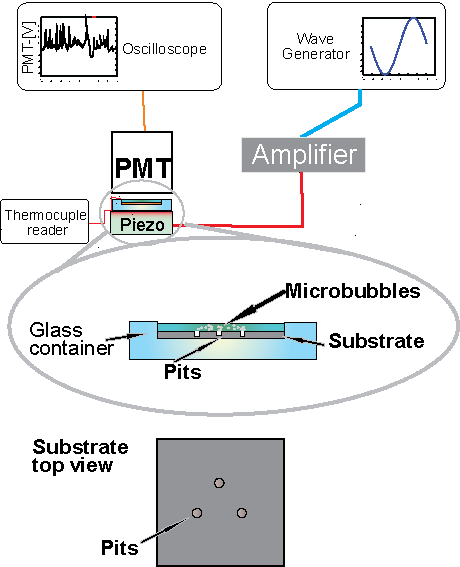}
\caption{Experimental setup (not to scale) showing the different components used. The zoomed inset shows the microreactor and a top view of a three-pit substrate.}
\label{fig:ExpSetup}
\end{figure}

A scheme of the experimental setup used is shown in Figure \ref{fig:ExpSetup}. The reaction chamber was a glass container of 25 mm outer diameter, 15 mm 
inner diameter and depth of 2 mm, and bottom thickness of 2 mm. The bottom 
thickness matched the quarter-wavelength vibration imparted by a
piezo Ferroperm PZ27 6 mm thick with a diameter of 25 mm, glued to the bottom 
of the reaction chamber. 

The ultrasonic wave was generated by a Hameg HM 8131-2 arbitrary 
waveform generator and amplified by a Krohn-Hite Model 7500 amplifier for the sonochemical reaction experiments and a LeCroy WaveSurfer 452 oscilloscope to read-out PMT measurements.

The powers used for the experiments were calculated from calorimetric measurements with a Hanna K-type Thermocouple leading to 12.7 W for the high power and 3.32 W for the low power settings.
Since control over the heating of the liquid volume was not available, temperature measurements were carried out before and after irradiation times of 3 minutes at similar conditions at which the SL and SCL signals were recorded. The thermocouple was removed from the liquid chamber during sonication to avoid damaging the thermocouple tip. At low power (3.32 W) the temperature did not increase by more than 3 K. For high power (12.7 W) the temperature increase was of 10 K. The 10 s period would generate an increase in temperature of around 0.6 K. We overestimate the temperature increase to be at least 1 K (2 K maximum) to be conservative.
For this reason, the PMT measurements lasted in general no longer than 10 s to avoid large temperature variations.

The continuously applied acoustic field generated a standing wave depending on the height of the liquid column. For liquid heights close to one-quarter or three-quarters of the acoustic wavelength (approximately  250 or 300 $\mu\ell$ volume of liquid, respectively) a pressure antinode is expected to be located on the substrate and a node at the free liquid-air interface. In this study frequencies of about 200 kHz with a corresponding water height of approximately 5 mm were used (250 $\mu\ell$).  

 A Hamamatsu E849-35 PMT (2.5 ns risetime), with 15 mm diameter glass window, amplified by a Canberra H.V. Supply Model 3002 was placed to capture the light emitted by the sonicated liquid on top of the chamber.

The experiments were conducted with different liquids but the same volumes (250 $\mu\ell$) at ambient conditions and open to the atmosphere. The voltage readout of the PMT corresponded to the SL and SCL emission, where applicable, from a certain population of bubbles. 

The variations to the bottom surface (square silicon substrate of 10 mm width) of the micro-sonoreactor (the same as presented in \citep{fernandez2010}) were: blank with no pit (equivalent to a conventional US reactor of the bath type), one, two or three pits (small predefined cylindrical crevices on the silicon surface). The pits acted as nucleation sites for microbubble streamers that would otherwise not be present at the conditions studied. 

Two different power settings were chosen out of the three presented in \citep{fernandez2010}, corresponding to lower and higher power levels. The main reason to select these two settings is that they evidence a clear difference in the bubble pattern and the ultrasonic power being supplied to the whole system. Hence, we expect to see differences in the bubble populations capable of emitting light and producing radicals in all cases.

The controlled and localized acoustic microbubble generation can be sustained 
for at least several hours due to dissolved gas in the liquid transported into the pit by a process similar to rectified diffusion 		
\citep{bre02,apfel1970role,crum1982nucleation,fowlkes1988cavitation}. Since temperature and gas escaping the microchamber could not be controlled in this particular case, all experiments were carried out within 5 to 10 minutes.

For the SCL imaging, a digital SLR camera (Nikon D90) with 18-55 mm AFS zoom lens using the settings ISO1250, f 5.0 and a 60 s exposure time.

For SL imaging a NF Multifunction synthesizer WF-1946A with a NF HSA-4014 amplifier, and a fan to cool the microreactor were used to get similar conditions as described before. The exposure time for experimental imaging and dark conditions subtraction was 10 minutes with a BitranBS-41L cooled CCD camera coupled to a Nikkor 35 mm lens and a magnifying glass.

\section{Results and Discussion}   \label{sect:Results}

\begin{figure}
\centering
\includegraphics[width=0.65\columnwidth]{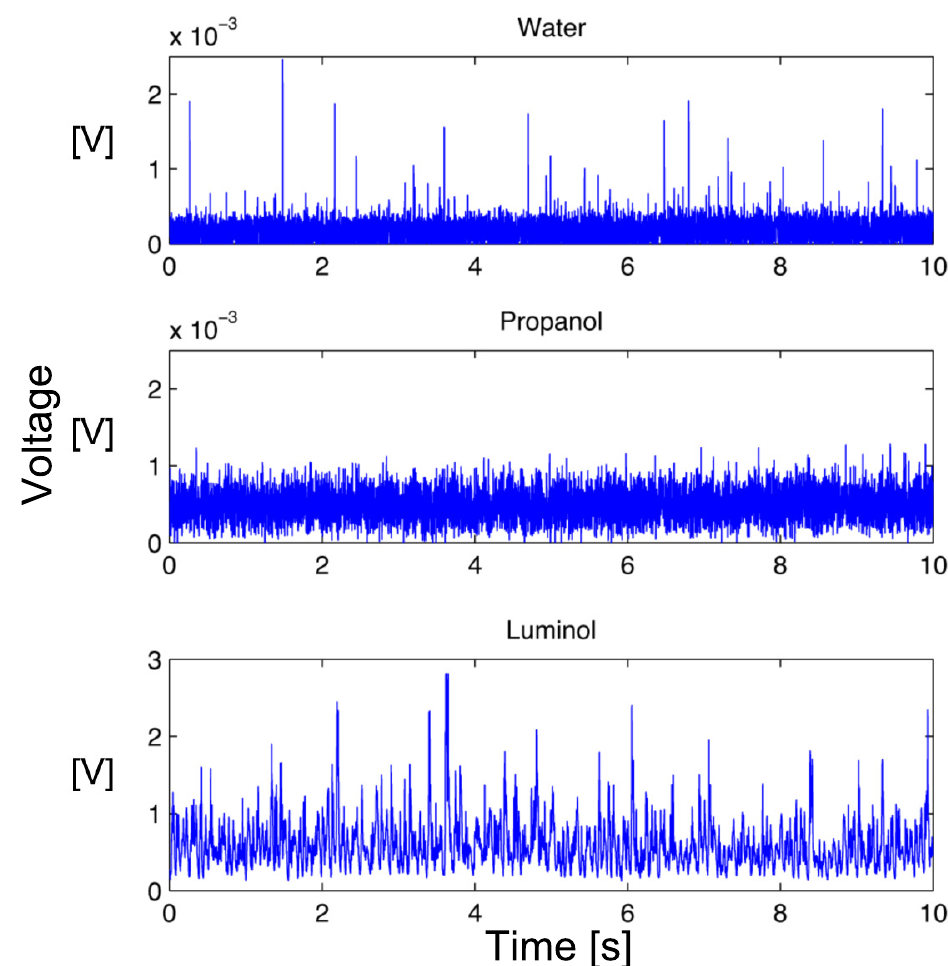}
\caption{Typical emission profiles recorded with the PMT for the different studied experimental conditions.}
\label{fig:typSpec}
\end{figure}

Contrary to the highly reproducible characteristics of single bubble cavitation, multiple bubbles are difficult to characterize since the bubble size distribution is constantly changing and bubbles do not cavitate always in phase with the driving frequency (period doubling and chaotic behaviors are reported in the literature \citep{lauterborn1981,cabeza1998}). However, the overall multiple bubble activity can be quantified by measuring the total SL intensity. Figure \ref{fig:typSpec} shows PMT output recorded for 10 s for the three systems studied. Despite the appearance of emission spikes (common for these systems \citep{ohl2011sonochemistry,negishi1961experimental}), the average intensity of each system was used for comparison in the following discussion. 

The emission spikes along a constant low intensity emission are short pulses that originate from specific bubbles that, upon reaching an appropriate size in the expansion phase, collapse and emit a strong light pulse. Micro-shocks occurring within the bubble during final collapse stages are reported elsewhere both for water and luminol solutions \citep{jarman1960,negishi1961experimental}.

It can be seen in Figure \ref{fig:typSpec} that the emission spikes mainly appear for the water and luminol system and not that frequently for the aqueous propanol system. 
Despite the less frequent presence of these spikes in the propanol system, the average intensity was higher than that observed for water. This behavior provides evidence that the bubble population is largely transient as there is no SL quenching \citep{lee2005effect,price2004sonoluminescence,toegel2002suppressing}. The presence of propanol can increase the bubble population by lowering the surface tension and facilitaing the pinch-off events of microbubbles from the pit. 
The white space in between the signal and the ``x-axis'' for propanol and luminol, not present in plain water, is due to a higher overall SL and SCL intensity.
A more detailed analysis for each system is presented in the following sections.

\subsection{Water}

\begin{figure}
	\centering
		\includegraphics[width=0.7\columnwidth]{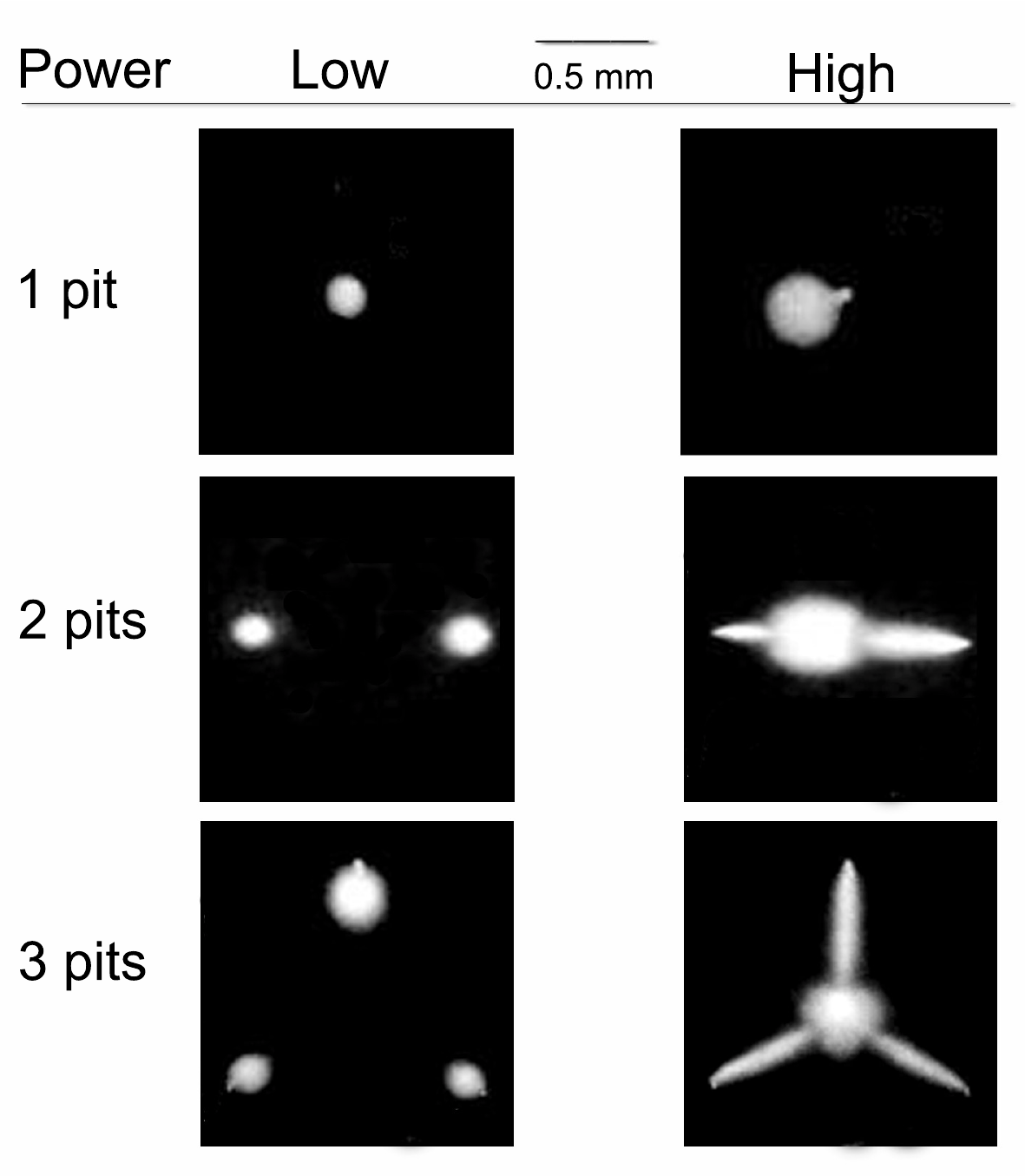}
		\caption{Top view of the visible bubble streamers for the different scenarios studied. See Supporting videos}
	\label{fig:visicomb}
\end{figure}

When water is poured over the silicon substrate and US is switched on, a cloud of bubbles appears from the micropits at low power levels.
An interesting behavior is observed when there are multiple pits driven at high power: the ejected bubbles travel to a common center point (see Figure \ref{fig:visicomb} and supporting videos) due to a complex interplay of primary and secondary Bjerknes forces \citep{fernandez2010}. 

\begin{figure}
\centering
\includegraphics[width=0.7\columnwidth]{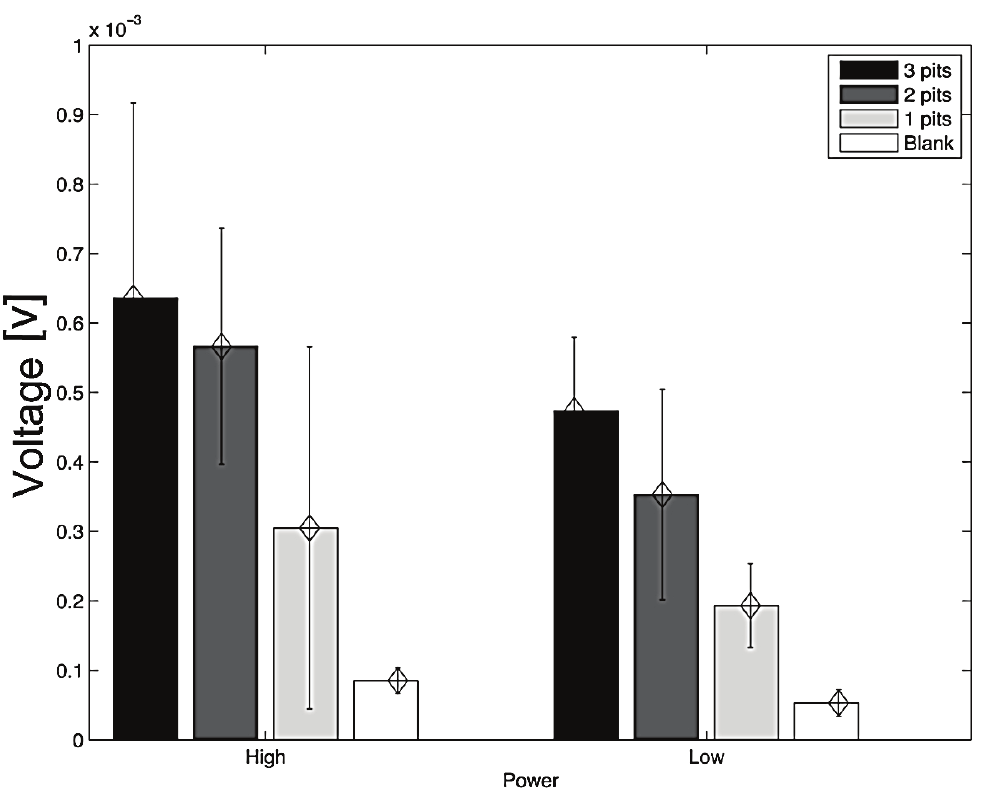}
\caption{Averaged SL intensities for the different studied experimental conditions in water.}
\label{fig:AIntW}
\end{figure}

Figure \ref{fig:AIntW} shows a clear trend of increasing SL intensity with increasing number of pits, both at the low and high power regimes. 

As the number of pits is increased, the number of SL active bubbles increases.
The increase is almost double in the case of one-pit compared with the blank substrate.
For two-pits, the SL enhancement is tripled, whilst for three-pits the SL is almost four times that of the blank substrate at high power.
While a similar increase is observed at the higher power level, the relative
increase with three-pits compared to two-pits is low. This might be due to stronger bubble clustering effects which are known to reduce the maximum expansion radius and shorten the bubble collapse duration \citep{yasui2008strongly}. 

\begin{figure}
	\centering
		\includegraphics[width=0.7\columnwidth]{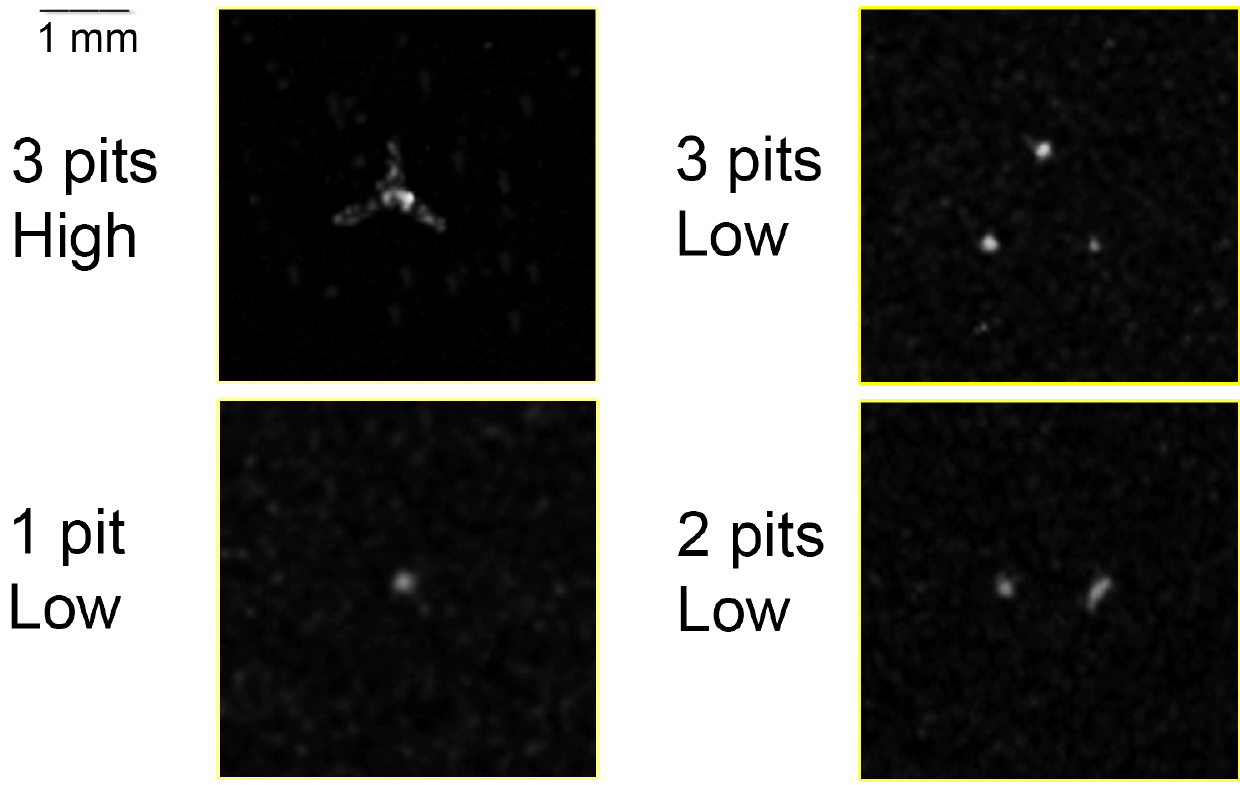}
		\caption{SL image from bubble streamers for different scenarios. The top row corresponds to three-pit cases with similar pattern as observed under visible conditions for high and low power. The remaining figures correspond to one-, two-pit cases at low power}
	\label{fig:SLcomb}
\end{figure}

When trying to image SL in the experiments with air-saturated Milli-Q water and the conditions described up to this point, not enough signal-to-noise images could be obtained. For this reason Milli-Q water was saturated with argon (Ar) and a glass slide was placed on top of the microreactor to avoid evaporation (see supporting video). During the imaging period the cooling of the piezo was carried out with a fan. The long exposure times required to obtain images like the ones presented in Figure \ref{fig:SLcomb} made it very difficult to cover the same experimental conditions as for the rest of this work. 

Nevertheless, these results allow us to conclude that the SL signal detected with the PMT was primarily due to light emission from the bubbles ejected from the micropits and not from random cavitation events in the bulk liquid.

\subsection{Aqueous propanol solution}

\begin{figure}
\centering
\includegraphics[width=0.7\columnwidth]{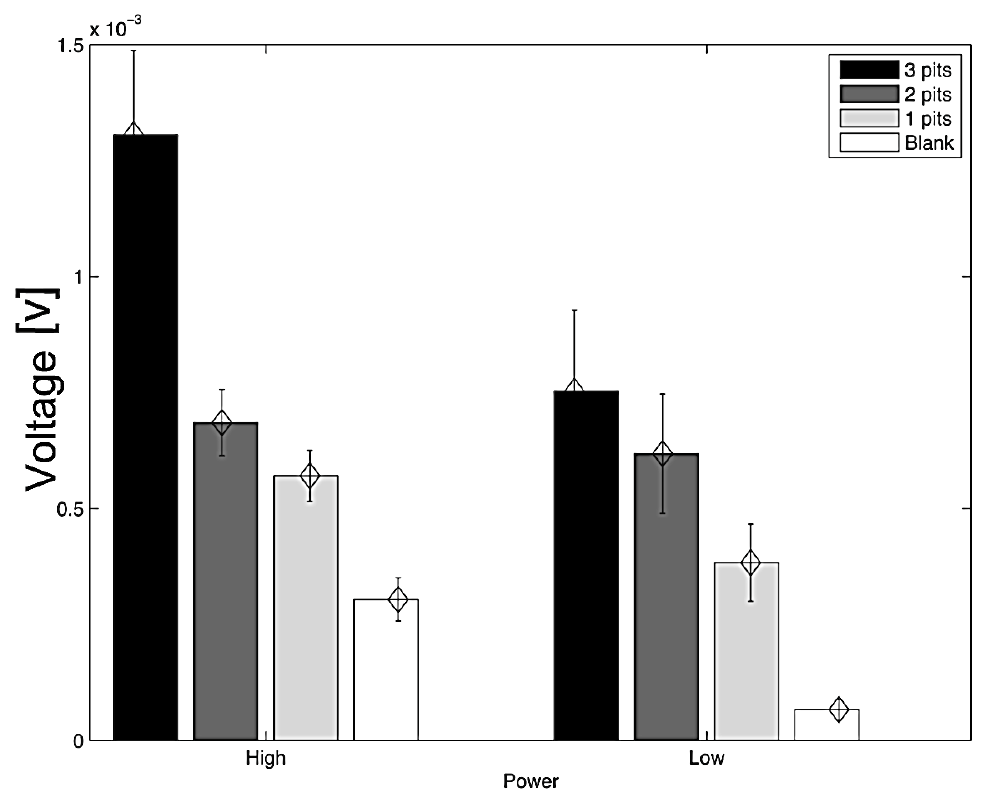}
\caption{Averaged SL intensities for the different studied experimental conditions with aqueous propanol solutions.}
\label{fig:JIntP}
\end{figure}

The visible bubble pattern for aqueous propanol solutions had no evident change when compared to water; however, as presented in Figure \ref{fig:typSpec}, spikes present in the water emission profile are not as frequent in propanol solutions and the spike height is on average lower.
The average intensities for propanol solutions are presented in Figure \ref{fig:JIntP}.

Propanol can cause two effects in sonicated liquids: by adsorbing to the bubble solution interface it hinders bubble coalescence. And due to its volatile nature, it evaporates into the bubble and lowers the polytropic exponent, resulting in less heating inside the bubble \citep{toegel2002suppressing}. These two effects can affect the SL in two ways. The hindrance to bubble coalescence has been shown to increase the number of transient cavitation bubbles \citep{price2004sonoluminescence}. The volatile nature leads to significant SL quenching in stable cavitation bubbles \citep{guan2003time}. 
However, both processes need sufficient time for the propanol to accumulate at the bubble interface. Here, under the conditions of transient cavitation this time is not given.
The observation that the average SL intensity for the propanol system instead of quenching the SL signal is higher than that observed in water is in agreement with the fact that the cavitation bubbles generated in the microreactor are transient in nature. As suggested earlier, the lower surface tension of the bubble stabilized on the pit due to the presence of propanol might yield a higher number of bubble streamers generated that in turn contribute to the increase in SL signal. This needs to be supported by future experiments with fast imaging of all these conditions.

\subsection{Luminol solution}

\begin{figure}
	\centering
		\includegraphics[width=0.7\columnwidth]{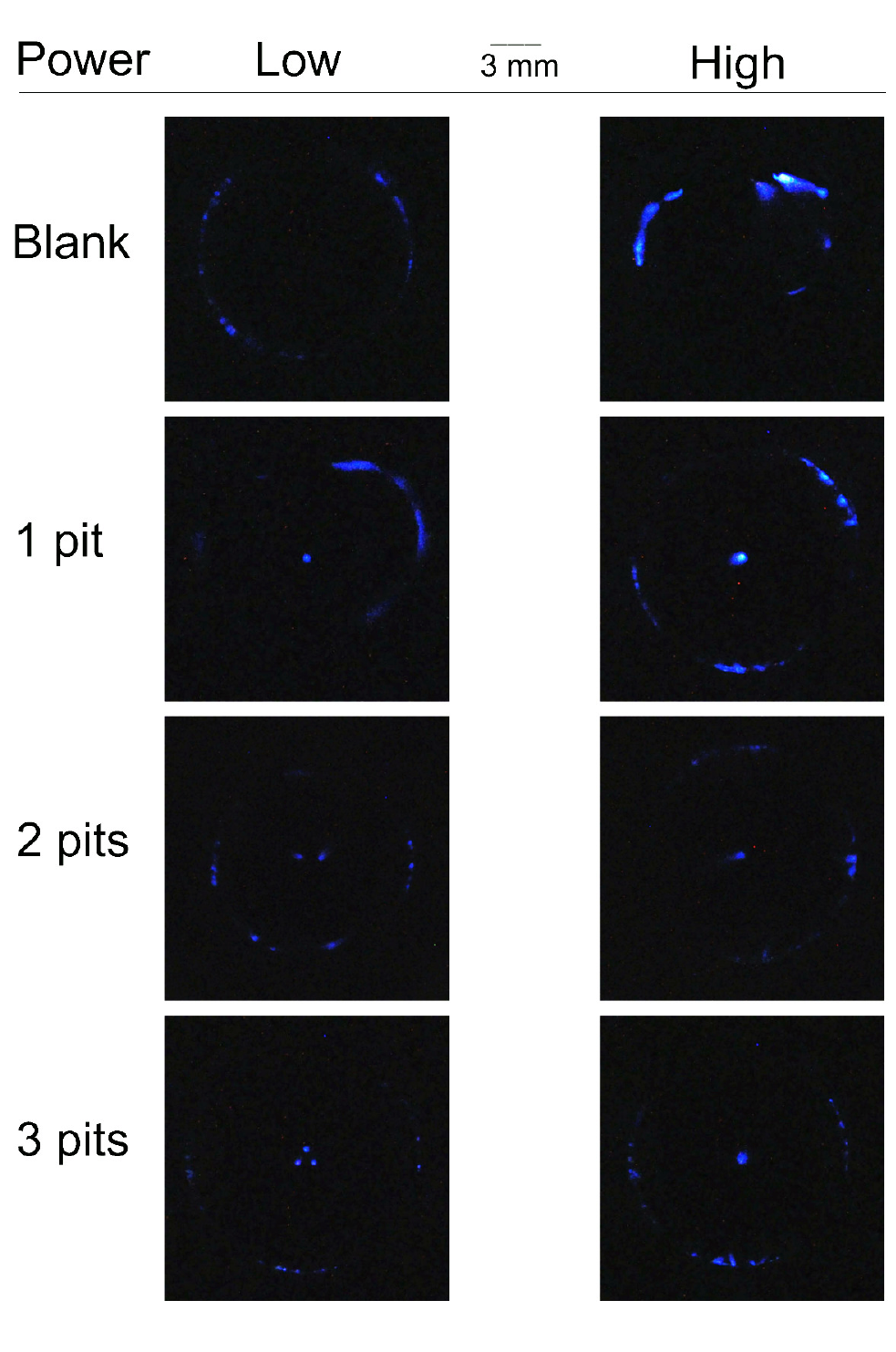} 
		\caption{Luminol solution images showing sonochemical active regions in white for the different scenarios studied.}
	\label{fig:lumcomb}
\end{figure}

Similar bubble streamer patterns from the luminol emission photographs are obtained (Figure \ref{fig:lumcomb}) to those observed in Figure \ref{fig:visicomb}. The light emitted was bright enough to be seen with the naked eye in adjusted dark conditions.

\begin{figure}
\centering
\includegraphics[width=0.7\columnwidth]{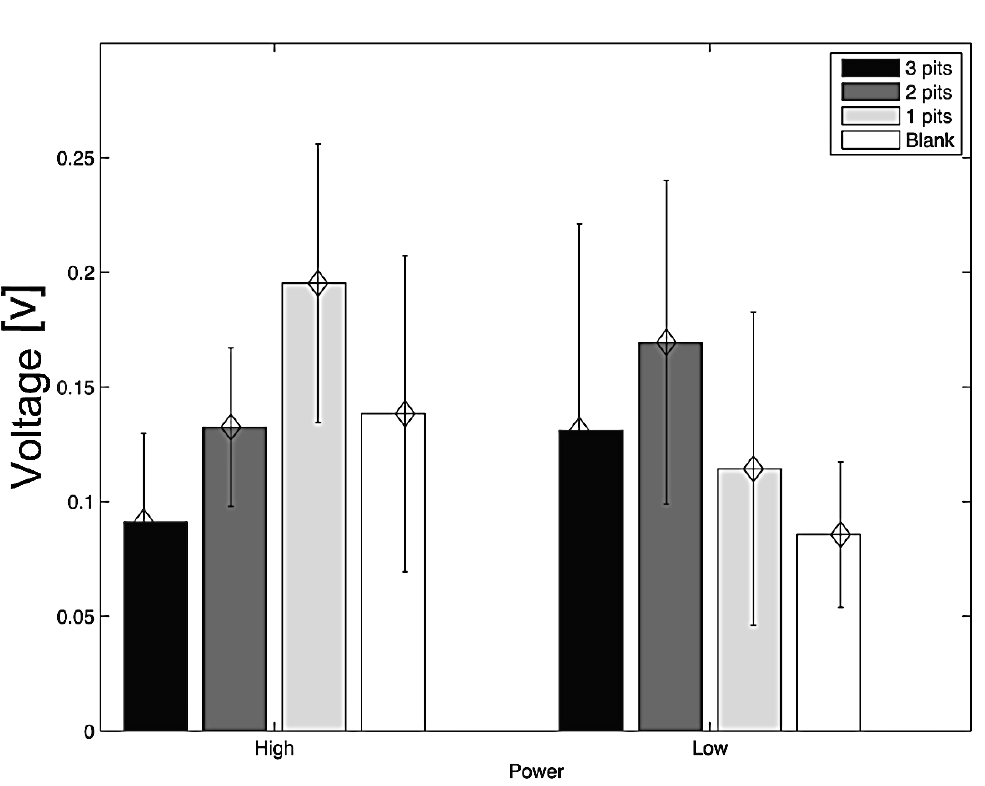}
\caption{Averaged SCL intensities for the different experimental conditions for aqueous luminol solution.}
\label{fig:CIntL}
\end{figure}

As can be seen in these images, the SCL and hence radical formation is intensified at the location of the micropits. The average SCL intensities for the case of luminol solutions were several orders of magnitude higher than those from the SL in water as presented in Figure \ref{fig:CIntL}. It is surprising then that the total emissions arising from the substrates containing pits are more or less identical to those arising from the blank substrate at low power.  
For the higher power, the two and three pits systems actually produce a lower yield than the blank substrate. It has been reported before that luminol SCL can have a unusual dependence with increasing power, sometimes reaching saturation and complete fading of intensity \citep{negishi1961experimental}.

An explanation for our observations is that by driving the systems with two or more pits at high power, the shape of the bubbles in the cluster become deformed. This, combined with liquid flow inside the microreactor chamber, mixing, free liquid-air interface oscillation and temperature increase in less than 5 min can bring a change in the SCL.
 
We speculate that the observation of different SCL signal trends when compared to our previous study measuring OH$^{.}$ radicals \citep{fernandez2010}  are due to the surface oscillation taking place at the liquid-air interface of the microreactor at the higher power conditions as reported before  \citep{tuziuti2010influence}. More details will be provided in the coming section \ref{subsect:CompSLSCL} and also in the supplementary videos.

\begin{figure}
\centering
\includegraphics[width=0.6\columnwidth]{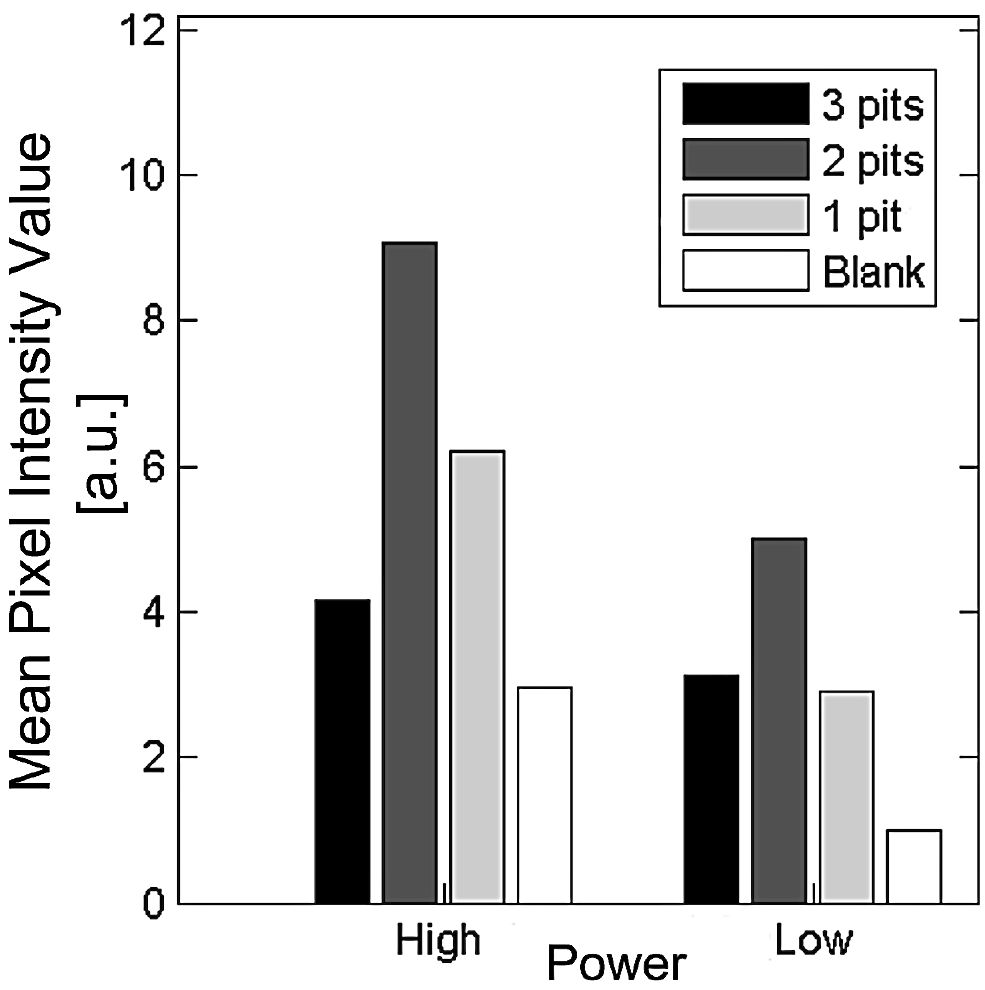}
\caption{Averaged pixel intensity values from the luminol images  recorded (SCL) for the different experimental conditions for aqueous luminol solutions. Besides the blue light coming from the center of the silicon substrate (microbubbles generated from the pits) in some cases some SCL signal is seen coming from the edges of the substrate; presumably from crevices existing in the glass reaction chamber or substrate edges.}
\label{fig:ImgSCL}
\end{figure}

The pixel intensities of the luminol photographs were also averaged in time and are presented in Figure \ref{fig:ImgSCL}. 
It can be seen that the two different experimental techniques (PMT and photographic imaging) produced similar trends (compare Figure \ref{fig:CIntL} and \ref{fig:ImgSCL}).
From the photographic images taken, it can be seen that luminol emission is intensified not just at the location of the pits, but also at the edges of the substrate and other locations like cracks on the silicon substrate edge or the corners of the microchamber which can act as nucleation sites. 
Such effects occur across all the experiments and will be additive to the luminol emission from the pits.
The PMT used to measure the SCL yield picks up light not just from the pits but also from the other places described.


\subsection{Comparing SL and SCL} \label{subsect:CompSLSCL}

\begin{figure}
\centering
\includegraphics[width=0.7\columnwidth]{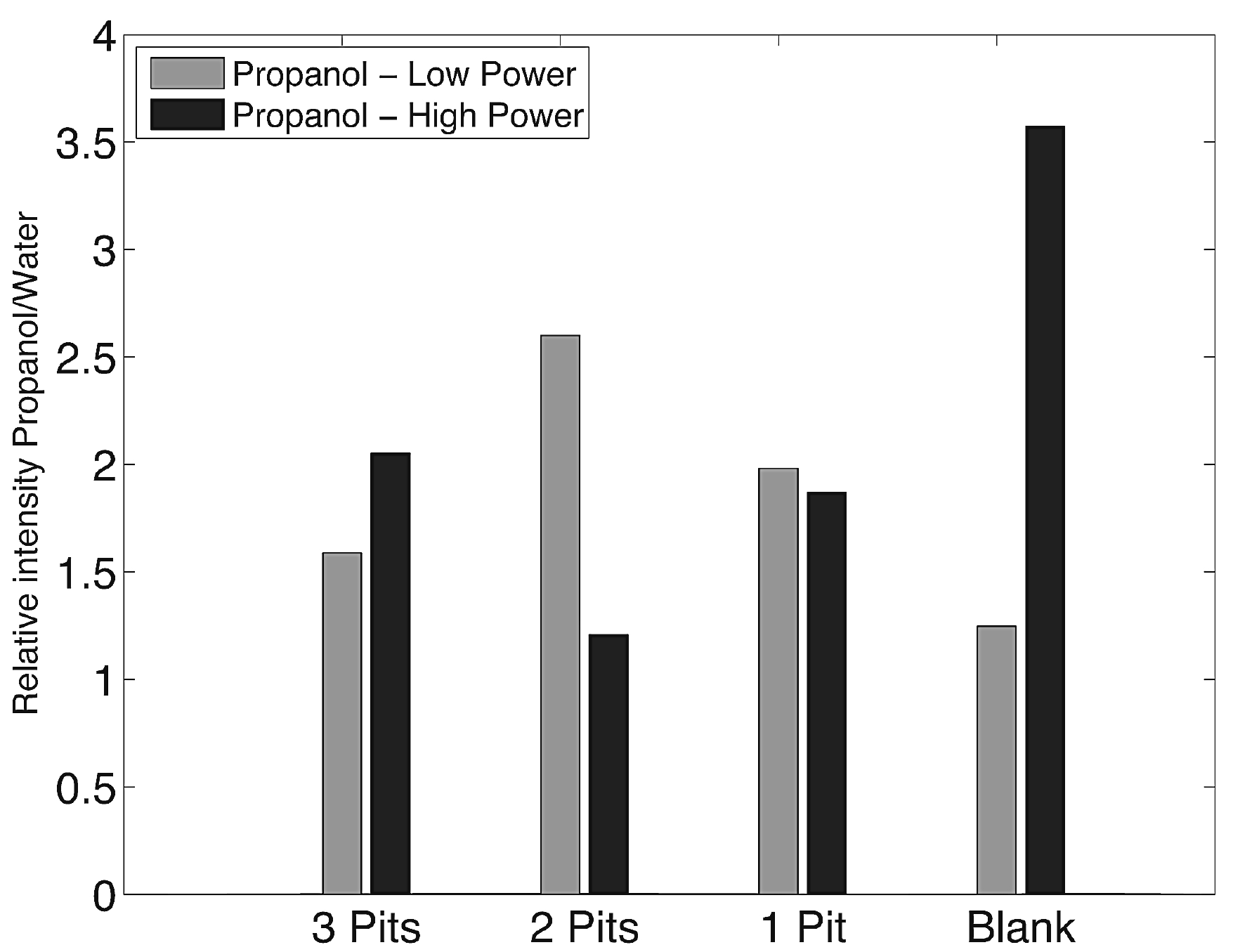}
\caption{Averaged SL intensities for the different experimental conditions for propanol solutions relative to water.}
\label{fig:Prop-wat-comp}
\end{figure}

In Figure \ref{fig:Prop-wat-comp} the relative increase of the SL intensity in water and propanol solutions is compared.
Note that for all situations there is an increase in SL intensity of propanol over water as described before. For two- and one-pit cases the low power shows a higher relative value than at high power (when the bubble streamers travel parallel to the wall and towards the center point) where the bubble population vary as the bubble clouds change their shape. 

\begin{figure}
\centering
\includegraphics[width=0.7\columnwidth]{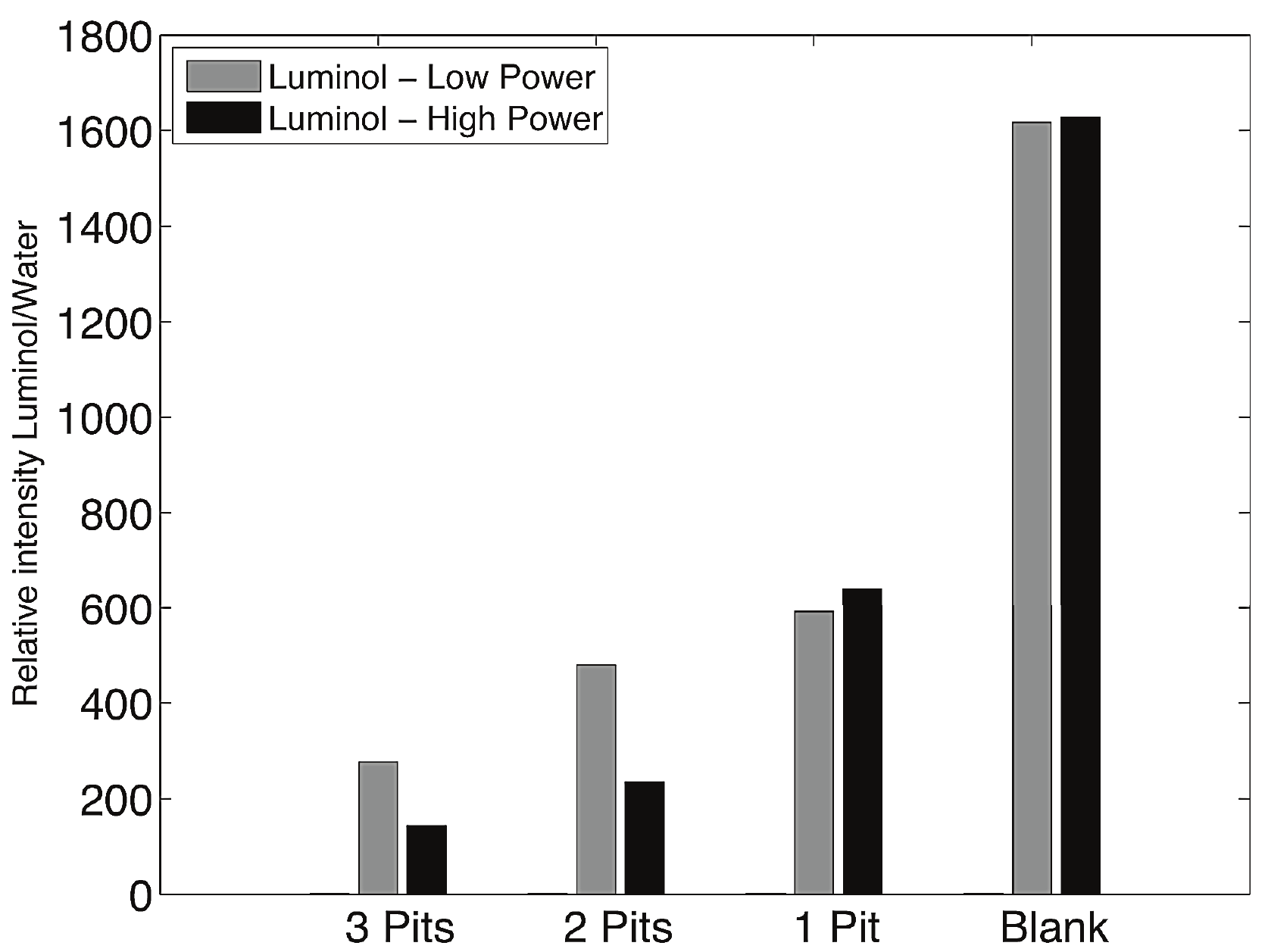}
\caption{Averaged SCL and SL intensities for the different experimental conditions for luminol solutions relative to water.}
\label{fig:Lum-wat-comp}
\end{figure}

When comparing the relative increase in SCL intensity of luminol over the SL signal from water, Figure \ref{fig:Lum-wat-comp} is obtained. Interestingly again the relative intensity is higher for the low power than for high power (now for the two- and three-pit systems) in line with our previous findings that higher power can be detrimental to OH$^{.}$ radical formation \citep{fernandez2010}. This can be linked with the fact that at low power the population of smaller and more spherical bubbles is larger, and it is expected that these are the ones contributing the most to radical production. Another possibility according to numerical simulations is that OH$^{.}$ production rate decreases at a too high temperature inside an air bubble as OH$^{.}$ is consumed by oxidizing nitrogen inside a bubble \citep{yasui2004optimum}.ÊThis may suggest that on average, the temperature inside the bubbles is higher with pits compared to the blank conditions.
Additionally, the bubble temperature is sometimes increased by the bubble-bubble interaction with smaller bubbles \citep{yasui:3233}. ÊIn the situation with pits, smaller and larger bubbles may result in higher bubble temperature inside larger bubbles compared to the case of no pit (blank) when there are mainly tiny bubbles (their size is too small to be recorded by the cameras).

Comparing Figures \ref{fig:Prop-wat-comp} and \ref{fig:Lum-wat-comp} it can be seen that the blank case is a clear indication of the strong influence of the contribution from the bubbles generated at the pits. For the case of one pit it can be observed that there is almost no change in the relative intensity value for both the propanol and luminol when increasing power. This strongly evidences that the interaction of the microbubble streamers generated by more than one pit is an important factor in the trends observed (cluster-cluster interactions through shockwave emission). 

To partially illustrate the complexity of this system the influence of  bubble-bubble and bubble-boundary interactions on the observed SL should be considered.
When microbubbles are exposed to an acoustic field, if smaller than resonant size, they tend to travel to the pressure antinodes and cluster by virtue of Bjerknes forces. This behavior may influence the generated SL. An example is the case of high power for the two- and three-pit systems, where there is a change in the microbubble pattern when compared to the one-pit and blank configuration.
A bubble cloud reflects and absorbs the sound field such that a lower intensity  will be experienced by bubbles inside the cloud due to shielding. This reduces the intensity of the bubble collapse and the active bubble population, leading to a reduced SL intensity when compared to conventional multibubble cavitation \citep{zeravcic2011}. As mentioned above bubble collapse in clusters also results in smaller expansion maximum radius and shorter collapse times \citep{yasui2008strongly}. 
 
  It has been modeled and experimentally confirmed for a single bubble that the strength of the bubble collapse is affected by its translational movement (accelerated due to added mass forces while the driving pressure increases), and that the strength of the bubble collapse 
  and its sphericity (i.e., the focusing power) are key ingredients determining the SL and SCL intensity \citep{Brenner95,sadighi2009dependence,hatanaka2002single,bre02}. Indeed, in our experiments higher SL is observed for higher power. 
  
Referring back to Figure \ref{fig:visicomb}:
At high power, for the two- and three-pit cases (two last figures in the right column of Figure \ref{fig:visicomb}), we see bubble clustering in between the pits, leading to a {\it depletion} of bubbles directly above the respective pits.
Therefore, in these cases, the bubbles are actually only cavitating in a thin liquid layer of a width of about 200 $\mu$m above the surface \citep{fernandez2012ts}. Consequently,
   there is then less mutual shielding of the bubbles
   as compared to the other four cases  (one-pit  case at both powers and  two- and three-pit cases
      at low powers). Additionally, at higher power there will be bubbles that expand to a larger size than at lower power, resulting in an increase in SL. We also expect that shockwave emissions from bubble clusters and cluster-cluster interactions (among microbubble streamers generated at each pit) influence the overall SL and SCL in terms of bubble maximum radius and collapse time as has been demonstrated before \citep{yasui2008strongly}.

Observations at the highest power give evidence that water can occasionally splash out from the liquid-air interface due to acoustic radiation force (see supplementary video). This has been reported to decrease the efficiency of sonochemical reactions \citep{tuziuti2010influence}, consistent with the reduced SCL signal at high power and regimes with more than one pit. As the meniscus shakes vigorously (a sizzling sound accompanies this process), the cavitation field changes considerably in a way difficult to quantify or predict. This can introduce sources of errors in the SL and SCL measurements. In addition, at the highest power heating and degassing of the liquid occur faster and this obviously changes the conditions for SL and SCL.  

It would be interesting to correlate the observed SL profile measured in this study to the influence of fluid mechanics such as flow due to acoustic streaming, microstreaming by meniscus oscillation, microbubbles flowing at different regimes, the interplay of secondary Bjerknes forces and bubble cluster interaction with the overall flow. Other factors such as closing the system with the presence of a glass slide on top of the microreactor to minimize liquid splashing, are among several other conditions that could be investigated but the appearance of degassing bubbles on the glass surface might be an undesired side-effect (see supplementary video). The above mentioned factors would influence the light emission, both from SL and SCL, at the different powers. As a last effort, continuous recording of these bubble streamers (in water, propanol and luminol solutions) may provide further clues to better understand our findings.

\section{Conclusions}    \label{sect:Conclus}
We have measured light emission at a single frequency from different systems at two different power levels in the presence and absence of pits on surfaces.  These pits promoted nucleation of microbubble streamers with distinctive streaming patterns at each power level. SL and SCL imaging provided evidence that the microbubbles arising from the pits are responsible for most of the light emission detected with PMT measurements.
The presence of transient cavitation conditions was verified by measuring the SL intensity in propanol solutions.
A difference in the light intensity of SL and SCL also led to the conclusion that there is a difference in the bubble population able to emit light and those which are chemically active. The multiple factors affecting the bubble streamers behavior are difficult to resolve individually. For that reason, there exists significant scope for future studies, particularly involving the control of the free surface liquid-air meniscus.

\section*{Acknowledgements}  \label{sect:Ack}
We thank Prof. Franz Grieser and Prof. Andrea Prosperetti for useful discussions and Stefan Schlautmann for technical assistance. This research was supported by the Technology Foundation STW, Applied Science Division of NWO and the Technology Programme of the Ministry of Economic Affairs, The Netherlands. 
The authors acknowledge financial support from the ARC, Australia.

\appendix
 \section*{Silicon substrate micromachining} \label{sect:App-Si}
 
\begin{figure}
            \centering
                        \includegraphics[width=0.65\columnwidth]{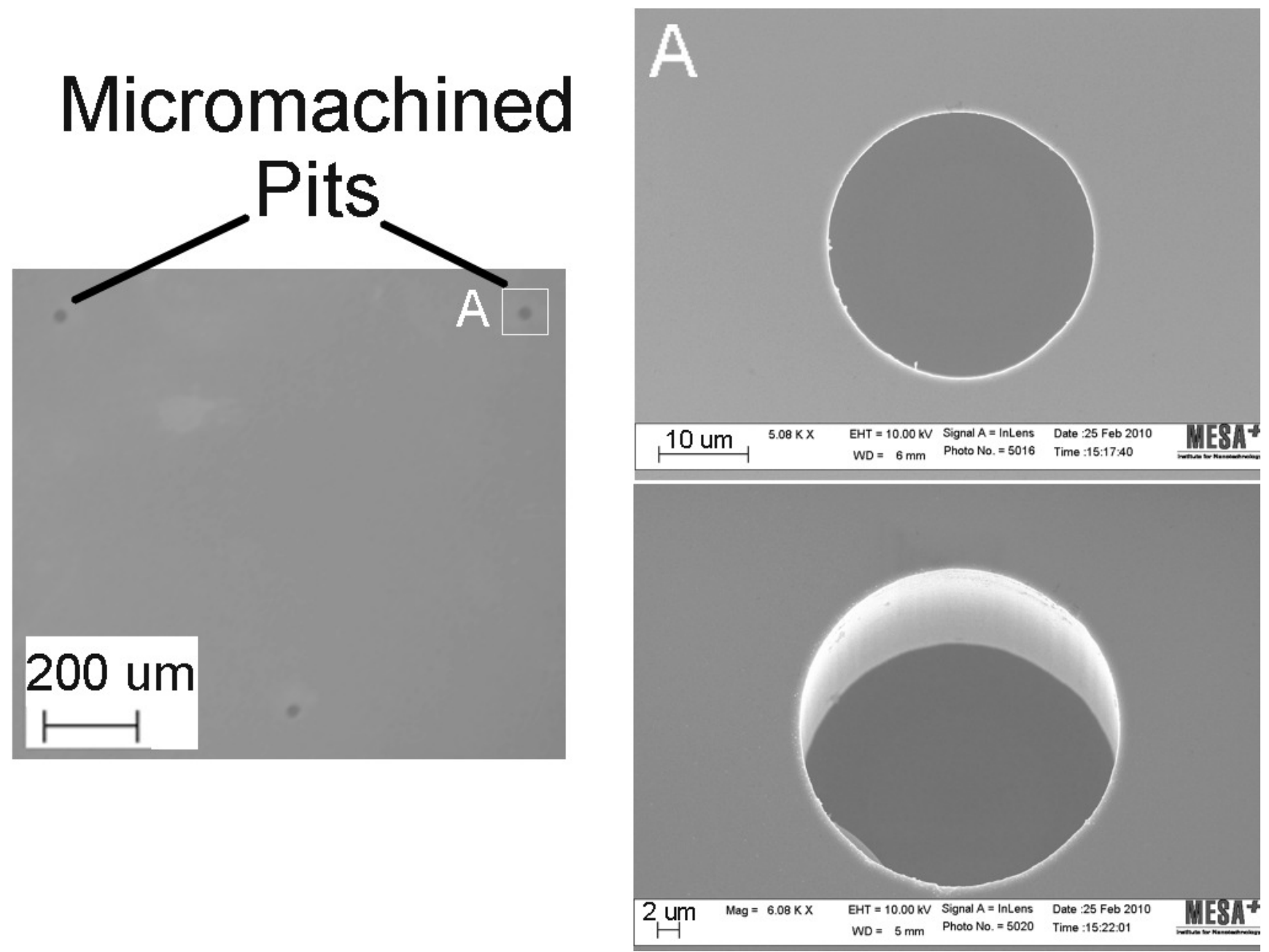}
            \caption{Pits etched on a silicone substrate. Top view, perspective view, top overview and distances.}
            \label{fig:PitsSi}
\end{figure}

 Three different configurations of pits were designed. The pits had the same diameter (30$\mu$m) and were arranged in sets of one, two (in a line) or three 
(in a triangle) at a distance of 1000 $\mu$m from each other (see Figure \ref{fig:PitsSi}).						
The substrates were micromachined under clean room conditions on double-side polished silicon wafers and spin coated with the photosensitive substance Olin 12, on which the designed pattern was transferred with a mask aligner EV620 (photolitography). After development the pit pattern was open and with a plasma dry-etching machine Adixen AMS 100 SE (Alcatel) process BHARS, the holes were etched into the silicon substrate at the desired depth. The machined diced silicon
 square pieces of 1 cm-side were mounted to the bottom of a small glass container which contained a liquid volume of 300 $\mu\ell$, to the bottom of which a piezo element was attached.			
 
 \section{PMT voltage levels} \label{sect:offoff}
\begin{figure}
\centering
\includegraphics[width=0.65\columnwidth]{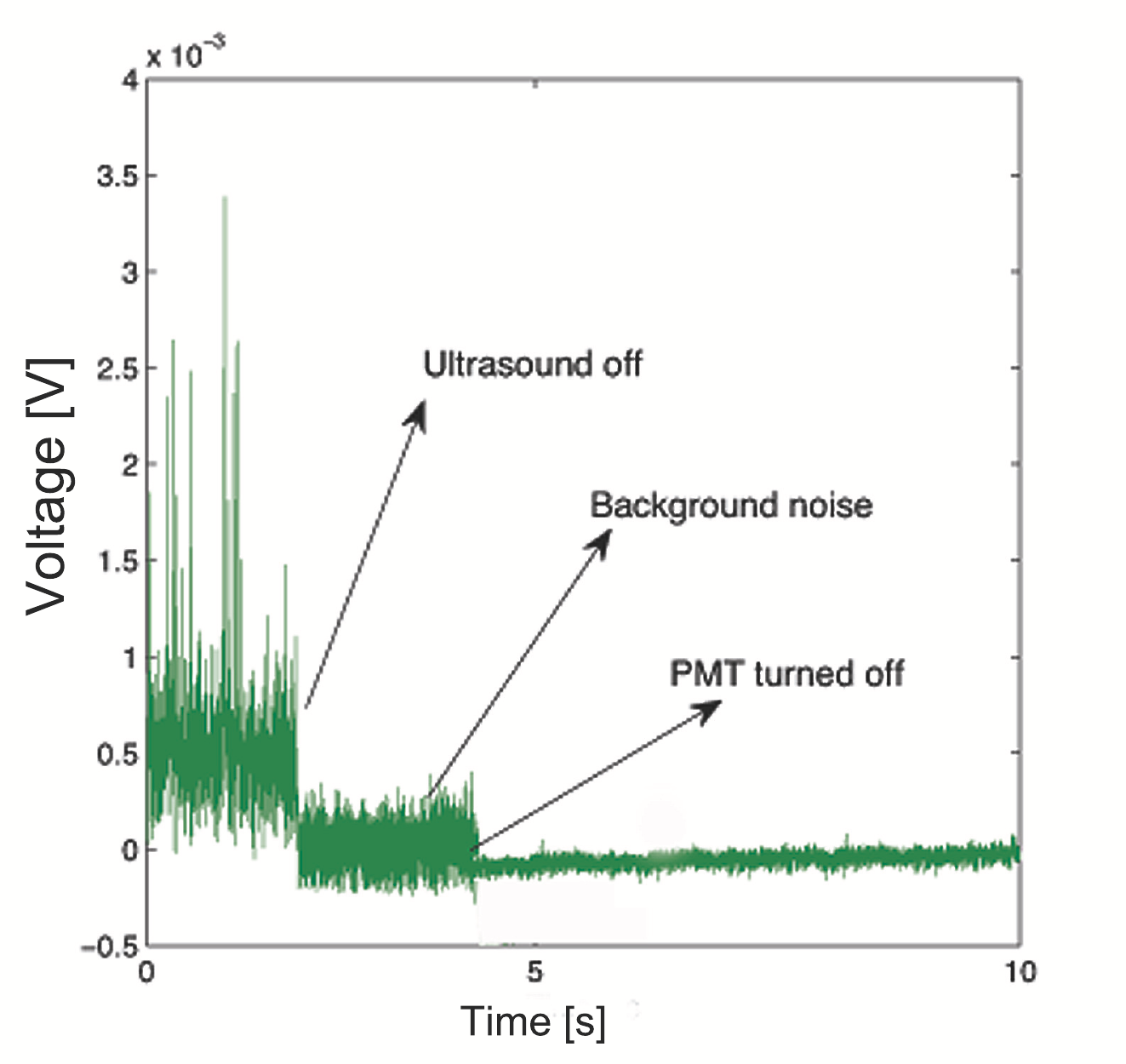}
\caption{PMT read-out case in which given instants are pointed by arrows. The arrows mark the instant where the US signal and PMT amplifier are turned off.}
\label{fig:offoff}
\end{figure}

 In Figure \ref{fig:offoff} the comparison of voltage read-out from the PMT when the ultrasound is turned off and when the PMT is turned off is shown.

%
%


\begin{thebibliography}{58}
\expandafter\ifx\csname natexlab\endcsname\relax\def\natexlab#1{#1}\fi

\bibitem[Apfel(1970)]{apfel1970role}
{\sc Apfel, R.~E.} 1970 Role of impurities in cavitation threshold
  determination. {\em J. Acoust. Soc. Am.\/} {\bf 48}~(5), 1179--1186.

\bibitem[Ashokkumar {\em et~al.\/}(2009)Ashokkumar, Lee, Iida, Yasui, Kozuka,
  Tuziuti \& Towata]{ashokkumar2009detection}
{\sc Ashokkumar, M., Lee, J., Iida, Y., Yasui, K., Kozuka, T., Tuziuti, T. \&
  Towata, A.} 2009 The detection and control of stable and transient acoustic
  cavitation bubbles. {\em Phys. Chem. Chem. Phys.\/} {\bf 11}~(43),
  10118--10121.

\bibitem[Ashokkumar {\em et~al.\/}(2010)Ashokkumar, Lee, Iida, Yasui, Kozuka,
  Tuziuti \& Towata]{AshokSpatial2010}
{\sc Ashokkumar, Muthupandian, Lee, Judy, Iida, Yasuo, Yasui, Kyuichi, Kozuka,
  Teruyuki, Tuziuti, Toru \& Towata, Atsuya} 2010 Spatial distribution of
  acoustic cavitation bubbles at different ultrasound frequencies. {\em
  ChemPhysChem\/} {\bf 11}~(8), 1680--1684.

\bibitem[Ashokkumar {\em et~al.\/}(2007)Ashokkumar, Lee, Kentish \&
  Grieser]{ashokkumar2007bubbles}
{\sc Ashokkumar, M., Lee, J., Kentish, S. \& Grieser, F.} 2007 Bubbles in an
  acoustic field: an overview. {\em Ultrason. Sonochem.,\/} {\bf 14}~(4),
  470--475.

\bibitem[Borkent {\em et~al.\/}(2009)Borkent, Gekle, Prosperetti \&
  Lohse]{borkent2009nucleation}
{\sc Borkent, Bram~M., Gekle, Stephan, Prosperetti, Andrea \& Lohse, Detlef}
  2009 Nucleation threshold and deactivation mechanisms of nanoscopic
  cavitation nuclei. {\em Phys. Fluids\/} {\bf 21}~(10), 102003.

\bibitem[Bremond {\em et~al.\/}(2006{\natexlab{{\em a\/}}})Bremond, Arora,
  Dammer \& Lohse]{bremond2006interaction}
{\sc Bremond, Nicolas, Arora, Manish, Dammer, Stephan~M. \& Lohse, Detlef}
  2006{\natexlab{{\em a\/}}} Interaction of cavitation bubbles on a wall. {\em
  Phys. Fluids\/} {\bf 18}~(12), 121505.

\bibitem[Bremond {\em et~al.\/}(2006{\natexlab{{\em b\/}}})Bremond, Arora, Ohl
  \& Lohse]{bre06}
{\sc Bremond, N., Arora, M., Ohl, C.~D. \& Lohse, D.} 2006{\natexlab{{\em
  b\/}}} Controlled multi-bubble surface cavitation. {\em Phys. Rev. Lett.\/}
  {\bf 96}, 224501.

\bibitem[Brenner {\em et~al.\/}(2002)Brenner, Hilgenfeldt \& Lohse]{bre02}
{\sc Brenner, M.~P., Hilgenfeldt, S. \& Lohse, D.} 2002 Single bubble
  sonoluminescence. {\em Rev. Mod. Phys.\/} {\bf 74}, 425--484.

\bibitem[Brenner {\em et~al.\/}(1995)Brenner, Lohse \& Dupont]{Brenner95}
{\sc Brenner, Michael~P., Lohse, Detlef \& Dupont, T.~F.} 1995 Bubble shape
  oscillations and the onset of sonoluminescence. {\em Phys. Rev. Lett.\/} {\bf
  75}, 954--957.

\bibitem[Brotchie {\em et~al.\/}(2009)Brotchie, Grieser \&
  Ashokkumar]{brotchie2009}
{\sc Brotchie, A., Grieser, F. \& Ashokkumar, M.} 2009 Effect of power and
  frequency on bubble-size distributions in acoustic cavitation. {\em Phys.
  Rev. Lett.\/} {\bf 102}~(8), 84302.

\bibitem[Cabeza {\em et~al.\/}(1998)Cabeza, Sicardi-Schifino, Negreira \&
  Montaldo]{cabeza1998}
{\sc Cabeza, C., Sicardi-Schifino, AC, Negreira, C. \& Montaldo, G.} 1998
  Experimental detection of a subharmonic route to chaos in acoustic cavitation
  through the tuning of a piezoelectric cavity. {\em J. Acoust. Soc. Am.\/}
  {\bf 103}, 3227.

\bibitem[Chen {\em et~al.\/}(2002)Chen, Matula \& Crum]{chen2002}
{\sc Chen, W.S., Matula, T.J. \& Crum, L.A.} 2002 The disappearance of
  ultrasound contrast bubbles: observations of bubble dissolution and
  cavitation nucleation. {\em Ultrasound Med. Biol.\/} {\bf 28}~(6), 793--803.

\bibitem[Cobley {\em et~al.\/}(2011)Cobley, Edgar, Goosey, Kellner \&
  Mason]{cobley2011initial}
{\sc Cobley, A.J., Edgar, L., Goosey, M., Kellner, R. \& Mason, T.~J.} 2011
  Initial studies into the use of ultrasound to reduce process temperatures and
  chemical usage in the pcb desmear process. {\em Circuit World\/} {\bf
  1}~(37), 15--23.

\bibitem[Crum(1982)]{crum1982nucleation}
{\sc Crum, L.A.} 1982 {Nucleation and stabilization of microbubbles in
  liquids}. {\em Appl. Sci. Res.\/} {\bf 38}~(1), 101--115.

\bibitem[Crum(1994)]{crum1994}
{\sc Crum, LA} 1994 Sonoluminescence, sonochemistry, and sonophysics. {\em J.
  Acoust. Soc. Am.\/} {\bf 95}, 559.

\bibitem[Didenko \& Suslick(2002)]{Didenko2002}
{\sc Didenko, Y.~T. \& Suslick, K.~S.} 2002 The energy efficiency of formation
  of photons, radicals and ions during single-bubble cavitation. {\em Nature\/}
  {\bf 418}~(6896), 394--397.

\bibitem[Dijkink {\em et~al.\/}(2008)Dijkink, Le~Gac, Nijhuis, van~den Berg,
  Vermes, Poot \& Ohl]{Dijkink2008}
{\sc Dijkink, R., Le~Gac, S., Nijhuis, E., van~den Berg, A., Vermes, A., Poot,
  A. \& Ohl, C.~D.} 2008 Controlled cavitation-cell interaction: trans-membrane
  transport and viability studies. {\em Phys. Med. Biol.\/} {\bf 2}~(53), 375.

\bibitem[Fernandez~Rivas {\em et~al.\/}(2010)Fernandez~Rivas, Prosperetti,
  Zijlstra, Lohse \& Gardeniers]{fernandez2010}
{\sc Fernandez~Rivas, David, Prosperetti, Andrea, Zijlstra, Aaldert~G., Lohse,
  Detlef \& Gardeniers, Han J. G.~E.} 2010 Efficient sonochemistry through
  microbubbles generated with micromachined surfaces. {\em Angew. Chem. Int.
  Ed\/} {\bf 49}~(50), 9699--9701.

\bibitem[Fernandez~Rivas {\em et~al.\/}(2012)Fernandez~Rivas, Stricker,
  Zijlstra, Gardeniers, Lohse \& Prosperetti]{fernandez2012ts}
{\sc Fernandez~Rivas, David, Stricker, L, Zijlstra, A~G, Gardeniers, H J G~E,
  Lohse, D \& Prosperetti, A} 2012 Ultrasound artificially nucleated bubbles
  and their sonochemical radical production. {\em Ultrason. Sonochem.\/} {\bf
  Accepted}.

\bibitem[Flint \& Suslick(1991)]{flint1991}
{\sc Flint, E.B. \& Suslick, K.S.} 1991 The temperature of cavitation. {\em
  Science\/} {\bf 253}~(5026), 1397.

\bibitem[Fowlkes \& Crum(1988)]{fowlkes1988cavitation}
{\sc Fowlkes, J.~B. \& Crum, L.~A.} 1988 Cavitation threshold measurements for
  microsecond length pulses of ultrasound. {\em J. Acoust. Soc. Am.\/} {\bf
  83}~(6), 2190--2201.

\bibitem[Frenzel \& Schultes(1934)]{frenzel34}
{\sc Frenzel, H. \& Schultes, H.} 1934 Luminescenz im ultraschallbeschickten
  wasser. {\em Z. Phys. Chem.\/} {\bf B2742}~(421).

\bibitem[G.~Cravotto(2012)]{Cravotto2012}
{\sc G.~Cravotto, P.~Cintas} 2012 Harnessing mechanochemical effects with
  ultrasound-induced reactions. {\em Chem. Sci.\/} {\bf 2}~(3), 295--307.

\bibitem[Guan \& Matula(2003)]{guan2003time}
{\sc Guan, J. \& Matula, T.J.} 2003 Time scales for quenching single-bubble
  sonoluminescence in the presence of alcohols. {\em J. Phys. Chem. B\/} {\bf
  107}~(34), 8917--8921.

\bibitem[Hatanaka {\em et~al.\/}(2002)Hatanaka, Mitome, Yasui \&
  Hayashi]{hatanaka2002single}
{\sc Hatanaka, S., Mitome, H., Yasui, K. \& Hayashi, S.} 2002 Single-bubble
  sonochemiluminescence in aqueous luminol solutions. {\em J. Am. Chem. Soc.\/}
  {\bf 124}~(35), 10250--10251.

\bibitem[Hilgenfeldt {\em et~al.\/}(1999)Hilgenfeldt, Grossmann \&
  Lohse]{Hilgenfeldt99}
{\sc Hilgenfeldt, Sascha, Grossmann, Siegfried \& Lohse, Detlef} 1999 A simple
  explanation of light emission in sonoluminescence. {\em Nature\/} {\bf
  398}~(6726), 402--405.

\bibitem[Iida {\em et~al.\/}(2004)Iida, Yasui, Tuziuti, Sivakumar \&
  Endo]{iida2004ultrasonic}
{\sc Iida, Y., Yasui, K., Tuziuti, T., Sivakumar, M. \& Endo, Y.} 2004
  Ultrasonic cavitation in microspace. {\em Chem. Commun.\/} {\bf 20},
  2280--2281.

\bibitem[Jarman(1960)]{jarman1960}
{\sc Jarman, P.} 1960 Sonoluminescence: a discussion. {\em J. Acoust. Soc.
  Am.\/} {\bf 32}, 1459.

\bibitem[Labouret \& Frohly(2002)]{labouret2002}
{\sc Labouret, S. \& Frohly, J.} 2002 Bubble size distribution estimation via
  void rate dissipation in gas saturated liquid. application to ultrasonic
  cavitation bubble fields. {\em Eur. Phys. J. Appl. Phys.\/} {\bf 19}~(01),
  39--54.

\bibitem[Lauterborn \& Cramer(1981)]{lauterborn1981}
{\sc Lauterborn, W. \& Cramer, E.} 1981 Subharmonic route to chaos observed in
  acoustics. {\em Phys. Rev. Lett.\/} {\bf 47}~(20), 1445--1448.

\bibitem[Lauterborn \& Kurz(2010)]{lauterborn2010}
{\sc Lauterborn, W. \& Kurz, T.} 2010 Physics of bubble oscillations. {\em Rep.
  Prog. Phys.\/} {\bf 73}, 106501.

\bibitem[Lee {\em et~al.\/}(2005)Lee, Kentish \& Ashokkumar]{lee2005effect}
{\sc Lee, J., Kentish, S.E. \& Ashokkumar, M.} 2005 The effect of
  surface-active solutes on bubble coalescence in the presence of ultrasound.
  {\em J. Phys. Chem. B\/} {\bf 109}~(11), 5095--5099.

\bibitem[Lee {\em et~al.\/}(2007)Lee, Tuziuti, Yasui, Kentish, Grieser,
  Ashokkumar \& Iida]{lee2007influence}
{\sc Lee, J., Tuziuti, T., Yasui, K., Kentish, S., Grieser, F., Ashokkumar, M.
  \& Iida, Y.} 2007 Influence of surface-active solutes on the coalescence,
  clustering, and fragmentation of acoustic bubbles confined in a microspace.
  {\em J. Phys Chem. C\/} {\bf 111}~(51), 19015--19023.

\bibitem[Lohse(2005)]{lohse2005sonoluminescence}
{\sc Lohse, D.} 2005 {Sonoluminescence-Cavitation hots up}. {\em Nature\/} {\bf
  434}, 33--34.

\bibitem[Luther {\em et~al.\/}(2001)Luther, Mettin, Koch \&
  Lauterborn]{luther2001}
{\sc Luther, S., Mettin, R., Koch, P. \& Lauterborn, W.} 2001 Observation of
  acoustic cavitation bubbles at 2250 frames per second. {\em Ultrason.
  Sonochem.\/} {\bf 8}~(3), 159--162.

\bibitem[Marinesco \& Trillat(1933)]{marinesco1933}
{\sc Marinesco, M. \& Trillat, J.J.} 1933 Action des ultrasons sur les plaques
  photographiques. {\em Compt. Rend.\/} {\bf 196}, 858--860.

\bibitem[Marmottant {\em et~al.\/}(2006)Marmottant, Raven, Gardeniers, Bomer \&
  Hilgenfeldt]{marmottant2006microfluidics}
{\sc Marmottant, P., Raven, J.~P., Gardeniers, H., Bomer, J.~G. \& Hilgenfeldt,
  S.} 2006 Microfluidics with ultrasound-driven bubbles. {\em J. Fluid Mech.\/}
  {\bf 568}, 109--118.

\bibitem[Matula \& Roy(1997)]{matula1997}
{\sc Matula, T.J. \& Roy, R.A.} 1997 Comparisons of sonoluminescence from
  single-bubbles and cavitation fields: bridging the gap. {\em Ultrason.
  Sonochem.\/} {\bf 4}~(2), 61--64.

\bibitem[Negishi(1961)]{negishi1961experimental}
{\sc Negishi, Katsuo} 1961 Experimental studies on sonoluminescence and
  ultrasonic cavitation. {\em J. Phys. Soc. Jpn.\/} {\bf 16}~(7), 1450--1465.

\bibitem[N.V.~Dezhkunov(2004)]{Dezhkunov2004}
{\sc N.V.~Dezhkunov, T.G.~Leighton} 2004 Study into correlation between
  ultrasonic capillary effect and sonoluminescence. {\em J. Eng. Phys.
  Thermophys.\/} {\bf 77}, 53--51.

\bibitem[N.V.~Dezhkunov(2005)]{Dezhkunov2005}
{\sc N.V.~Dezhkunov, V.~Dallacasa} 2005 Physical backgrounds for application of
  power ultrasound in fluorescent dye penetrant inspection. In {\em Proc. of
  4$^{th}$ European Congress on Acoustics\/}, pp. L231--L234. Budapest.

\bibitem[Ohl \& Wolfrum(2003)]{Ohl2003131}
{\sc Ohl, C.-D. \& Wolfrum, B.} 2003 Detachment and sonoporation of adherent
  hela-cells by shock wave-induced cavitation. {\em Bba-Gen Subjects\/} {\bf
  1-3}~(1624), 131--138.

\bibitem[Price {\em et~al.\/}(2004)Price, Ashokkumar \&
  Grieser]{price2004sonoluminescence}
{\sc Price, G.J., Ashokkumar, M. \& Grieser, F.} 2004 Sonoluminescence
  quenching of organic compounds in aqueous solution: Frequency effects and
  implications for sonochemistry. {\em J. Am. Chem. Soc.,\/} {\bf 126}~(9),
  2755--2762.

\bibitem[Sadighi-Bonabi {\em et~al.\/}(2009)Sadighi-Bonabi, Rezaei-Nasirabad \&
  Galavani]{sadighi2009dependence}
{\sc Sadighi-Bonabi, R., Rezaei-Nasirabad, R. \& Galavani, Z.} 2009 The
  dependence of the moving sonoluminescing bubble trajectory on the driving
  pressure. {\em J. Acoust. Soc. Am.\/} {\bf 126}, 2266.

\bibitem[Sunartio {\em et~al.\/}(2007)Sunartio, Yasui, Tuziuti, Kozuka, Iida,
  Ashokkumar \& Grieser]{sunartio2007}
{\sc Sunartio, D., Yasui, K., Tuziuti, T., Kozuka, T., Iida, Y., Ashokkumar, M.
  \& Grieser, F.} 2007 Correlation between {{Na*}} emission and ``chemically
  active'' acoustic cavitation bubbles. {\em ChemPhysChem\/} {\bf 8}~(16),
  2331--2335.

\bibitem[Suslick {\em et~al.\/}(1999)Suslick, Didenko, Fang, Hyeon, Kolbeck,
  McNamara, Mdleleni \& Wong]{suslick1999acoustic}
{\sc Suslick, K.S., Didenko, Y., Fang, M.M., Hyeon, T., Kolbeck, K.J.,
  McNamara, W.B., Mdleleni, M.M. \& Wong, M.} 1999 Acoustic cavitation and its
  chemical consequences. {\em Philos. T. R. Soc. A\/} {\bf 357}~(1751), 335.

\bibitem[Suslick \& Flannigan(2008)]{suslick2008inside}
{\sc Suslick, K.~S. \& Flannigan, D.~J.} 2008 Inside a collapsing bubble:
  Sonoluminescence and the conditions during cavitation. {\em Annu. Rev. Phys.
  Chem.\/} {\bf 59}, 659--683.

\bibitem[Tandiono {\em et~al.\/}(2011)Tandiono, Ohl, Ow, Klaseboer, Wong, Dumke
  \& Ohl]{ohl2011sonochemistry}
{\sc Tandiono, Ohl, S.W., Ow, D.S.W., Klaseboer, E., Wong, V.V., Dumke, R. \&
  Ohl, C.D.} 2011 Sonochemistry and sonoluminescence in microfluidics. {\em
  Proc. Natl. Acad. Sci. USA\/} {\bf 108}~(15), 5996--5998.

\bibitem[Toegel {\em et~al.\/}(2000)Toegel, Gompf, Pecha \&
  Lohse]{toegel2000does}
{\sc Toegel, R., Gompf, B., Pecha, R. \& Lohse, D.} 2000 Does water vapor
  prevent upscaling sonoluminescence? {\em Phys. Rev. Lett.\/} {\bf 85}~(15),
  3165--3168.

\bibitem[Toegel {\em et~al.\/}(2002)Toegel, Hilgenfeldt \&
  Lohse]{toegel2002suppressing}
{\sc Toegel, Ruediger, Hilgenfeldt, Sascha \& Lohse, Detlef} 2002 Suppressing
  dissociation in sonoluminescing bubbles: The effect of excluded volume. {\em
  Phys. Rev. Lett.\/} {\bf 88}, 034301.

\bibitem[Tsochatzidis {\em et~al.\/}(2001)Tsochatzidis, Guiraud, Wilhelm \&
  Delmas]{tsochatzidis2001}
{\sc Tsochatzidis, NA, Guiraud, P., Wilhelm, AM \& Delmas, H.} 2001
  Determination of velocity, size and concentration of ultrasonic cavitation
  bubbles by the phase-doppler technique. {\em Chem. Eng. Sci.\/} {\bf 56}~(5),
  1831--1840.

\bibitem[Tuziuti {\em et~al.\/}(2010)Tuziuti, Yasui, Kozuka \&
  Towata]{tuziuti2010influence}
{\sc Tuziuti, T., Yasui, K., Kozuka, T. \& Towata, A.} 2010 Influence of
  liquid-surface vibration on sonochemiluminescence intensity. {\em J. Phys.
  Chem. A\/} {\bf 114}~(27), 7321--7325.

\bibitem[W.B.~McNamara~III(2000)]{mcnamara00}
{\sc W.B.~McNamara~III, Y.~Didenko, K.S.~Suslick} 2000 Effect of noble gases on
  sonoluminescence temperatures during multibubble cavitation. {\em Phys. Rev.
  Lett.\/} {\bf 84}, 777--780.

\bibitem[Yasui {\em et~al.\/}(2008{\natexlab{{\em a\/}}})Yasui, Iida, Tuziuti,
  Kozuka \& Towata]{yasui2008strongly}
{\sc Yasui, K., Iida, Y., Tuziuti, T., Kozuka, T. \& Towata, A.}
  2008{\natexlab{{\em a\/}}} Strongly interacting bubbles under an ultrasonic
  horn. {\em Phys. Rev. E\/} {\bf 77}~(1), 016609.

\bibitem[Yasui {\em et~al.\/}(2011)Yasui, Towata, Tuziuti, Kozuka \&
  Kato]{yasui:3233}
{\sc Yasui, Kyuichi, Towata, Atsuya, Tuziuti, Toru, Kozuka, Teruyuki \& Kato,
  Kazumi} 2011 Effect of static pressure on acoustic energy radiated by
  cavitation bubbles in viscous liquids under ultrasound. {\em J. Acoust. Soc.
  Am.\/} {\bf 130}~(5), 3233--3242.

\bibitem[Yasui {\em et~al.\/}(2004)Yasui, Tuziuti \& Iida]{yasui2004optimum}
{\sc Yasui, K., Tuziuti, T. \& Iida, Y.} 2004 Optimum bubble temperature for
  the sonochemical production of oxidants. {\em Ultrasonics\/} {\bf 42}~(1),
  579--584.

\bibitem[Yasui {\em et~al.\/}(2008{\natexlab{{\em b\/}}})Yasui, Tuziuti, Lee,
  Kozuka, Towata \& Iida]{yasui2008}
{\sc Yasui, K., Tuziuti, T., Lee, J., Kozuka, T., Towata, A. \& Iida, Y.}
  2008{\natexlab{{\em b\/}}} The range of ambient radius for an active bubble
  in sonoluminescence and sonochemical reactions. {\em J. Chem. Phys.\/} {\bf
  128}, 184705.

\bibitem[Zeravcic {\em et~al.\/}(2011)Zeravcic, Lohse \& van
  Saarloos]{zeravcic2011}
{\sc Zeravcic, Zorana, Lohse, Detlef \& van Saarloos, Wim} 2011 Collective
  oscillations in bubble clouds. {\em J. Fluid Mech.\/} {\bf 680}, 114--149.

\end{thebibliography}
\end{document}